\documentclass[11pt]{article}

\usepackage{graphicx, xcolor, amsmath, pifont, comment, layout, amsthm, amssymb}

\usepackage{hyperref} 
\hypersetup{ colorlinks=true, linkcolor=black, filecolor=black, urlcolor=cyan, }

\usepackage{ulem}

\newcommand{\revDeleted}[1]{}
\newenvironment{revTwoAdded}{}{}
\newcommand{\revTwoDeleted}[1]{}

\newcommand{\revThreeDeleted}[1]{}

\begin{document}

\title{Is Ethereum Proof of Stake Sustainable?\\
--- Considering from the Perspective of Competition Among Smart Contract Platforms ---}
\author{Kenji Saito\footnote{Professor, Graduate School of Business and Finance, Waseda University},
Yutaka Soejima\footnote{Principal research fellow, SBI Financial and Economic Research Institute},
Toshihiko Sugiura\footnote{Principal research fellow, SBI Financial and Economic Research Institute \begin{revTwoAdded}(currently, Executive Officer, Rating and Investment Information, Inc.)\end{revTwoAdded}},
Yukinobu Kitamura\footnote{Professor, Faculty of Data Science, Rissho University;
Associated Research Scholar, Institute of Economic Research, Hitotsubashi University},\\
Mitsuru Iwamura\footnote{Emeritus Professor, Waseda University}}
\date{September 7, 2023}

\maketitle

\begin{abstract}
Since the Merge update upon which Ethereum transitioned to Proof of Stake,
it has been touted that it resulted in lower power consumption and increased security.
However, even if that is the case, can this state be sustained?

In this paper, we focus on the potential impact of competition with other smart contract
platforms on the price of Ethereum's native currency, Ether (ETH), thereby
raising questions about the safety and sustainability purportedly brought about by the
design of Proof of Stake.
\begin{description}
\item[Keywords:] Proof of Stake, Blockchain,
Cryptocurrency, Smart Contract, Competition
\item[JEL classification:] B31, E42, E51
\end{description}
\end{abstract}

\pagebreak

\section{Introduction}\label{sec-intro}
\subsection{Motivation}
Bitcoin\cite{Nakamoto2008:Bitcoin}, unlike traditional central bank payment systems,
established a currency creation and transfer mechanism based on a decentralized autonomous
system rooted in competition and collaboration among participants.
Furthermore, it became the first attempt to design and implement the tool for that purpose,
known as {\it blockchain}.

Ethereum\cite{Buterin2013:Ethereum}, designed as a subsequent blockchain, realized the ability
to store and call up to execute program codes that represent changes in any state (though
conditionally, as detailed in Section~\ref{sec-blockchain}).
This not only enabled monetary transfers, i.e., changes in balances, but also made it possible to
perform actions such as transferring rights represented by moving tokens.

Both applied anti-fraud measures known as Proof of Work (PoW).
The former devised and the latter extended\footnote{
GHOST (Greedy Heaviest Observed Sub-Tree), which will be discussed later in Section~\ref{sec-eth-pos}.
} {\it Nakamoto Consensus}\footnote{
This name originates from Satoshi Nakamoto, the pseudonymous developer who developed the Bitcoin algorithm.
}, which posits that the history with the highest accumulated PoW cost is the most canonical history.
So far, by this approach, they have succeeded in preventing fraud at a certain high level.

However, in PoW, a vast amount of electricity is consumed to support the computational competition
among {\it miners}, and this consumption has grown to rival the total power consumption of medium-sized
developed nations.
Given recent environmental concerns, this has been pointed out as a significant problem (the issue
of energy consumption).
Additionally, both blockchains can process an extremely limited number of transactions\footnote{
Processing that cannot be divided and is handled as a single unit, such as the transfer of rights and
associated payments, etc.
} within a certain timeframe (the scalability issue), and there has been criticism that execution takes
too much time.

In September 2022, Ethereum abandoned PoW, which had issues like energy consumption and scalability,
and transitioned to another method called Proof of Stake (PoS) (this transition is referred to as
{\it The Merge}).
The authors are interested in whether the post-transition Ethereum can continue to prevent fraud at
the same level as before.

\subsection{Goals}
The goals of this paper can be summarized in the following three points:
\begin{enumerate}
\item Introduce a perspective into the discussion of Ethereum's sustainability that considers
the market price level of its native token\footnote{
Tokens that are generated as a reward for the maintenance activities of the blockchain.
}, Ether (ETH), and the competition among smart contract execution platforms.
\item Model PoS in Ethereum (hereafter referred to as Ethereum PoS) to lay a foundation for
discussing its price and competition among platforms
\item Utilize the model for a discussion to consider cases where Ethereum PoS may not be
sustainable.
\end{enumerate}

\section{Background}
\subsection{Blockchain}\label{sec-blockchain}
\begin{figure}[h]
\begin{center}
\includegraphics[scale=0.4]{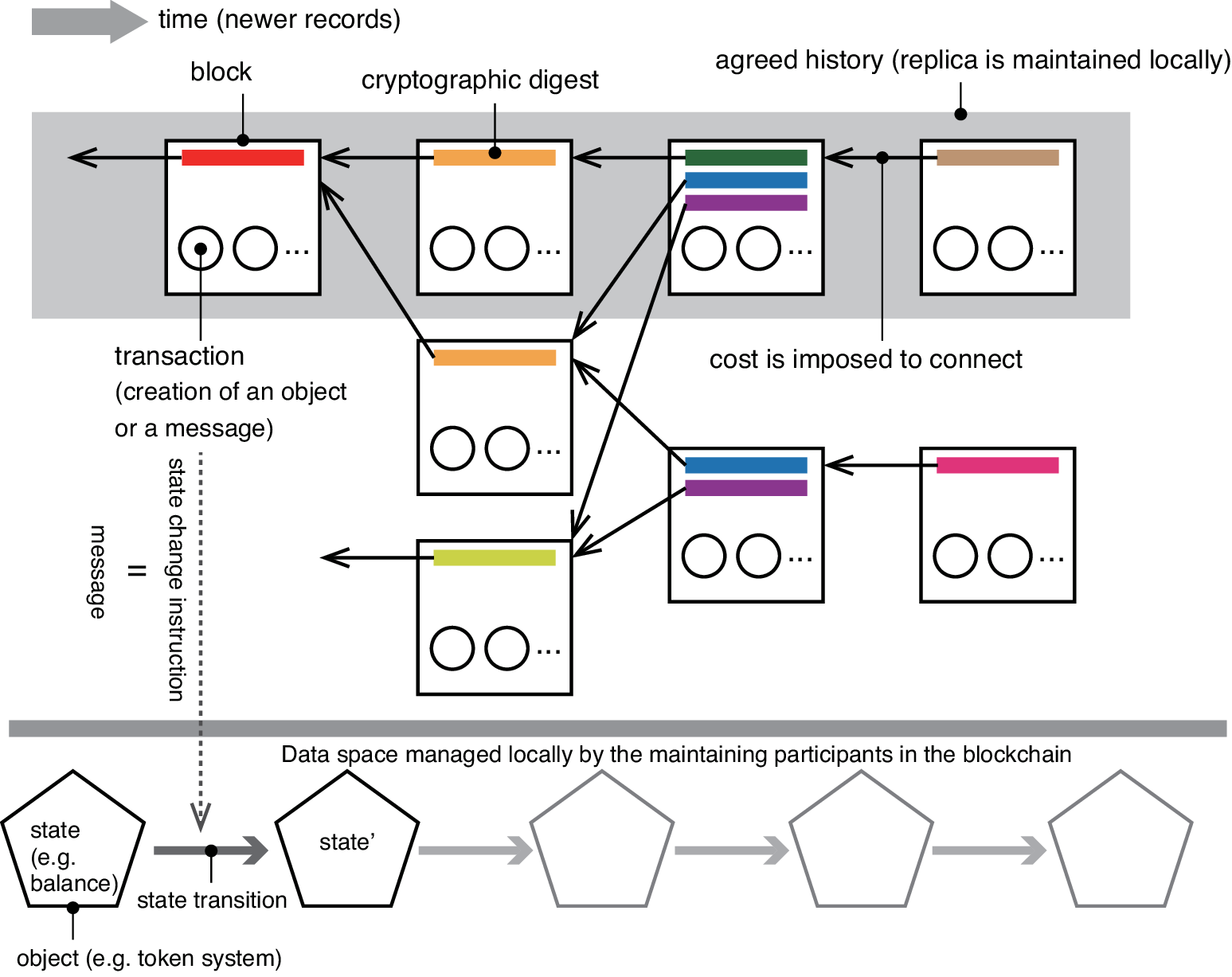}\\
{\footnotesize
\begin{itemize}
\item Each block in Bitcoin and Ethereum after The Merge has only a single cryptographic
digest that points to its preceding block.
\end{itemize}
}
\caption{Abstract Blockchain}\label{fig-abs-blockchain}
\end{center}
\end{figure}

Blockchains\cite{Nakamoto2008:Bitcoin} are generally designed as
illustrated in Figure~\ref{fig-abs-blockchain}.

According to a document\cite{Wood2015:Ethereum} that has been defining the technical
specifications of Ethereum from the beginning, a blockchain is described as a system that
manages the transition of states based on transactions ({\it state machine}).
That is, it maintains a set of key-value pairs ({\it variables}\footnote{When the value obtained
for a search key is always the same and does not change, it is specifically called a {\it constant}.
}), such as the balance of an account represented by a certain address, as its {\it state},
and by applying transactions, the system changes this state.

This state machine operates on a database (a data space managed locally) managed by
the participants maintaining the blockchain.
The blockchain's data structure itself is essentially a {\it ledger} that records transactions.

\paragraph{Smart Contracts}
The original motivation for creating such a ledger is believed to be to record the fact of
a transfer in an essentially unmodifiable form, so that no one could prevent one from freely
using the currency tokens they possess.

However, by recording program code, the data it handles (constants and variables), execution logs,
and changes in variables due to execution on such a ledger, it became possible for everyone to
verify that the state machine is working as intended (within the range written on the blockchain),
not just for monetary transfers but also for general state transitions.
This is what smart contracts\cite{Buterin2013:Ethereum} are, which were first fully introduced in
Ethereum.

The act of writing the program code of a smart contract and its initial data to the blockchain
is referred to as {\it deploying}.

\paragraph{State Machine Replication in Blockchain}
A block is a set of transactions (or rather, data that describe them).
Transactions cause changes in state by allowing the maintainers of the blockchain to
introduce new state machine objects (smart contracts) into the data space they manage locally,
send execution instructions (messages) with parameters to already introduced objects, or
by transferring native tokens.

In blockchain, participants voluntarily become validators, and each validator
verifies the validity\footnote{
For example, whether there is any inconsistency when compared with past records (such as
double-spending of tokens), or whether the transaction has been submitted by an authorized user
(such as if the current holder of the token has provided a valid signature when directing the transfer).
} of transactions they have chosen at their discretion.
A block is formed by collecting and recording the data of multiple verified
transactions\footnote{
Which validator can create a block will be discussed later in this section (in the case of PoW)
and in the next section (in the case of PoS).
}.

The execution of a blockchain is a process in which the state machine is fully replicated in
each participant's local data space autonomously, so that no one can stop the process.

The state machine replication, in the field of distributed systems, aims to achieve fault
tolerance by running multiple servers with the same functions (redundancy).
This method, which allows the system to continue despite the shutdown or malfunction of some
servers, was already established in the 1980s\cite{Schneider1990:SMR}.

For the state machine to be replicated, the following conditions must be met among the
participants performing the replication:
\begin{enumerate}
\item An identical initial state is shared.\label{smr-cond-init}
\item Transactions are shared in the same order.\label{smr-cond-order}
\item Transactions that have a non-deterministic\footnote{
A process that always returns the same output for the same input is called {\it deterministic},
and when the output is not necessarily the same, it is called {\it non-deterministic}.
In the state machine replication, non-deterministic state changes cannot be replicated, so
transactions cannot perform parallel processing or generate random numbers.\label{foot-deterministic}
} effect on the state are not applied.
\end{enumerate}

Therefore, in addition to sharing the first block (commonly known as the {\it genesis block}) as
mentioned above in \ref{smr-cond-init}, it is essential that all participants maintaining
the blockchain observe the sequence of transactions in the same order, as indicated in \ref{smr-cond-order}.
To achieve this, a cryptographic digest\footnote{The output value obtained by applying a
cryptographic hash function.
Cryptographic hash functions convert any input into a fixed bit-length value, such as 256 bits,
and always produce the same output for the same input.
However, if the input varies even slightly, an entirely different output is produced, and
it is unpredictable what that output will be until the computation is actually performed.
Furthermore, the output values are uniformly distributed across the space of numbers expressible
in that bit-length (for example, 256 bits), and the likelihood of two different inputs producing
the same output (a {\it collision}) is extremely low.
} of a block is stored in the block that follows it.
If block B stores the digest of block A, then there is an inherent requirement for block A
to precede block B.
This structure logically represents the chronology.
The arrows connecting the blocks in Figure~\ref{fig-abs-blockchain} are oriented in the opposite
direction of time progression because the digest stored in each block uniquely points to the
preceding block\footnote{
The possibility of a digest collision is extremely low.
Moreover, if a collision does occur and a digest identical to that of a past block is produced,
that new block can be invalidated.
Thus, blocks can be uniquely identified by their digests.
}.

However, in distributed systems in general, because participants might not always operate correctly,
or due to network transmission delays, the order of transactions may be observed differently.
In order to satisfy the condition mentioned in \ref{smr-cond-order} above, a mechanism for
consensus among participants, which uniquely determines the order of transactions, is necessary.

In blockchain as well, different blocks can be created simultaneously due to reasons such as
malicious participants or transmission delays, resulting in the branching ({\it forking}) of the
block sequence.
(In Figure~\ref{fig-abs-blockchain}, multiple branched histories are depicted, and since
validators form these while verifying all transactions in a block, there are no contradictions
within each history.)
Therefore, to satisfy the condition mentioned in \ref{smr-cond-order} above, when the block
sequence branches, a mechanism among participants is needed to agree on which sequence is
legitimate\footnote{
The term {\it legitimate} or {\it canonical} here does not mean that everyone follows some
predetermined legitimacy, but rather, it refers to a sequence that a substantial number of
participants recognize as the one that should be extended.
}.

In blockchain, a cost is imposed on the creation of blocks, ensuring that the cost accumulates
as blocks that had cost incurred to create them are linked together.
In other words, it is a mechanism where past blocks can only be tampered with by taking on
the accumulated costs anew.
By implementing this mechanism, the system aims to enhance tamper resistance.
In this context, among multiple branching sequences of blocks, the sequence with the greatest
accumulated cost is defined as the most legitimate history.
This ensures both the difficulty of tampering and the uniqueness of the history simultaneously.
This is known as {\it Nakamoto Consensus}.

\paragraph{Proof of Work and Proof of Stake}
The cost of creating a block can be rephrased as the conditions required to create a block.

In one method, a validator places a number ({\it nonce}: Number used Once) within the created
block, which they can decide arbitrarily.
The cryptographic digest of the block must be below a {\it target} value that is
typically inherited from the preceding block and adjusted by the same algorithm by all
participants.
This method of setting conditions is called Proof of Work (PoW).

In PoW, validators (miners) are compelled to repeatedly calculate the cryptographic hash
function, changing the nonce until they obtain a digest that meets the condition.
The act of receiving a {\it reward} (which will be mentioned later) based on the proof of
this work (that the block's digest is below the target) is called {\it mining}.

In PoW, the primary cost associated with the aforementioned computational work, in other words,
the power cost required for computer processing, constitutes the expense of creating a block.
As blocks are sequentially created and linked, the computational effort's cost (power cost)
accumulates, enhancing the system's resistance to tampering.
However, this method has been criticized for the vast amounts of power consumed due to the
extensive calculations carried out by a large number of miners.
Yet, as the importance of green energy grows and with the challenges of producing adjustable
power storage solutions that are cost-intensive, how to promote the adoption of such energy
has become a societal issue.
{\it Green mining}, which involves mining during times when there is an excess supply of
renewable energy, is gaining attention as a way to enhance the profitability of green energy
without relying on subsidies.
It is seen as a means to facilitate the transition towards a green economy\cite{Ibanez2023:green}.

Rather than such indirect methods, one alternative that is said to address these challenges
through algorithmic changes is Proof of Stake (PoS).
As we will elaborate later in section~\ref{sec-eth-pos}, this method involves
obtaining the right to vote on blocks and their sequences by depositing cryptographic assets,
with the legitimate history determined based on the votes received.

\paragraph{Rewards}
Participants involved in maintaining the legitimate block sequence (i.e., the agreed-upon official
history) receive native tokens, which the blockchain generates from nothing based on its protocol,
as compensation for their work.

\subsection{Proof of Work Economy}
If participants engage in maintaining the block sequence in pursuit of profits from the rewards
paid out in tokens, then the maintenance cost and the market value of the tokens will balance
out in the long term for the following reason.

As mentioned in the previous section, in the case of PoW, the primary cost of maintenance is
electricity.
When the token price is high and the expected value of the reward paid out in tokens exceeds the
electricity cost, new entries into mining occur.
With an increase in the overall computational power of participants, the frequency of obtaining
the appropriate block digest increases.
That is, the interval between block creations becomes shorter.
When the interval for block creation shortens, according to the protocol (as in the case of Bitcoin,
which aims to maintain a block creation interval of once every 10 minutes on average),
the {\it target} is adjusted to be smaller.
Consequently, the electricity cost required for a miner to earn one unit of token increases and
approaches the reward.
Conversely, when the token price is low, and the expected value of the reward falls below the
electricity cost, exits from mining occur (at least, the hardware is powered off), causing a
decrease in the overall computational power of the participants.
This results in a longer interval for block creation.
When the block creation interval lengthens, according to the protocol, the {\it target} is adjusted
to be larger.
As a result, the electricity cost required for a miner to earn one unit of token decreases,
again approaching the reward.

The authors discussed this in detail in \cite{Iwamura2019:BitcoinMonetaryPolicy}\cite{Saito2019:Stable}
and proposed design improvements to stabilize the price in PoW-based cryptocurrencies, which
tend to be volatile.
This paper attempts to apply a similar economic rationale to Ethereum PoS and analyze it.

\subsection{Ethereum Proof of Stake}\label{sec-eth-pos}
Ethereum originally adopted PoW, but transitioned to PoS after the completion of {\it The Merge} update
on September 15, 2022\cite{Ethereum2022:POS}\cite{Edgington2023:Eth2Book}.
In the following, we will explain the mechanism of PoS in Ethereum as a premise for understanding
the discussions in the subsequent sections.

\paragraph{Deposits}
Users who participate in maintaining the blockchain as validators need to pre-deposit the native
token of Ethereum, Ether (ETH) (of course, participants can choose not to become validators and
join Ethereum as users benefiting from transaction execution by paying fees).
The required deposit amount is 32 ETH per validator.
Rewards for participation are added to the deposit amount, and penalties (described later) are
deducted from it, so the deposit amount can increase or decrease after gaining participation
eligibility.
However, the effective deposit amount (effective balance\cite{McDonald2019:EffectiveBalance}) used
for calculating rewards and other purposes will never exceed 32 ETH.
Additionally, if the deposit amount falls below 16 ETH, one automatically loses the status of
a validator.

\paragraph{Slots and Epochs}
In Ethereum, time is divided into {\it slots} with 12-second intervals and {\it epochs} consisting
of 32 slots.

In each slot, a randomly selected validator creates at most one block (there may be cases where
a block is not created for reasons such as the selected validator being offline) and proposes it
to the whole network.
Then, a committee\footnote{
There are two types of committees responsible for attesting to blocks: the {\it beacon committee}
(named because, prior to The Merge, the PoS mechanism was operated on an independent {\it beacon chain}
separate from the PoW chain) and the {\it sync committee}.
The beacon committee is shuffled every epoch, and every validator is assigned to the committee for a slot.
The sync committee consists of 512 validators randomly chosen and is elected every 256 epochs.
The latter was introduced to facilitate block verification by clients that do not participate in the
blockchain maintenance activities ({\it light clients}).
} of validators, again randomly chosen, attests to the validity of the proposed block.
These attestations are signed, but attestations of the same content are aggregated using BLS
signatures\cite{Boneh2004:BLS}.

The first block of each epoch is called a {\it checkpoint}.
All validators attest to the validity of a pair of consecutive checkpoints $\langle c_{e-1}, c_e \rangle$
(where $e$ represents the number for the latest epoch).
Within this pair, the older one is referred to as the {\it source checkpoint}, and the newer one as
the {\it target checkpoint}.
For a pair that has gathered attestations equivalent to more than $\frac{2}{3}$ of the total deposit
amount of ETH by all validators:

\begin{enumerate}
\item $c_{e-1}$ (the source checkpoint) is {\it finalized}, and
\item $c_e$ (the target checkpoint) is {\it justified}.
\end{enumerate}

Since $c_{e-1}$ must have been justified\footnote{
During the attestation for the latest pair, it indicates the most recently justified checkpoint,
which defines the {\it source checkpoint}.
} when attestations were gathered for $\langle c_{e-2}, c_{e-1} \rangle$,
checkpoints undergo a phased certification of {\it justification} $\rightarrow$ {\it finalization}.
Typically, a block generated just before a checkpoint only takes about one slot's time to be
justified.
However, a block generated right after a checkpoint requires roughly the duration of one epoch to
be justified.
An additional epoch's duration is necessary for finalization.
Hence, if attestations proceed smoothly, transactions stored in a block will be finalized in less than
13 minutes (the time span of 2 epochs).

However, if enough attestations are not gathered, justification and finalization do not occur.
If there's a failure to finalize for five consecutive epochs, that is, if attestations amounting
to less than $\frac{2}{3}$ of the total deposit continue to be collected, there is a penalty that
confiscates from the deposits of validators who provided attestations different from the majority
or did not attest at all.
This confiscation is meant to encourage a situation where the majority eventually makes up greater than
or equal to $\frac{2}{3}$ of the total deposit, and finalization can take place.

\paragraph{Fork Choice}
Even in this mechanism, as is common with any distributed systems, discrepancies can occur due to
delays, failures, recovery from these issues, or even attacks, leading to the potential forking
of the sequence of blocks, or in other words, the chain.
In such cases, the total weight of attestations to blocks on each fork, converted based on their
represented effective balances, is compared after the justified checkpoint.
The branch with the highest total weight is adopted as the legitimate one according to the
LMD-GHOST\footnote{
LMD-GHOST stands for Last-Message Driven, Greedy Heaviest Observed SubTree.
GHOST is an extended Nakamoto consensus that was adopted in Ethereum.
} algorithm\cite{Buterin2020:LMDGHOST}.
This aspect, including the justification of checkpoints, constitutes the consensus\footnote{
In distributed systems, consensus indicates that the value of the same variable matches across all
concurrently running correct processes\cite{Lamport1998:Paxos}.
In the case of blockchain, the block number can be seen as the variable name, and the digest of
the block as its value.
} algorithm in the narrow sense in Ethereum's PoS system.

\paragraph{RANDAO}
As indicated in footnote~\ref{foot-deterministic}, if there are nondeterministic operations,
the state machine cannot be replicated.
Therefore, in the execution of a blockchain, it is not possible to obtain randomness in the
usual manner.
Instead, a deterministic and unpredictable method must be used to derive a random sequence
that everyone agrees upon.
Therefore, when it is necessary to make random selections, such as choosing the proposer of
a block or members of the beacon and sync committees, a random number called RANDAO (Random DAO)
is used.
This is a conceptual DAO (Decentralized Autonomous Organization)\cite{Buterin2014:DAO},
different from the commonly known ones realized through smart contracts, and is a 256-bit
value stored as part of Ethereum's state.
This value is shuffled as follows every time a block is proposed: the block's proposer signs
a data based on the epoch number with a verifiable BLS signature.
The new RANDAO value is the bitwise ``exclusive logical or'' (XOR) of the 256-bit digest of
that signature and the existing RANDAO value.

\paragraph{Gas Fee}
The computational resources required to execute a smart contract (such as the number of
computational steps or storage usage) are measured in units called {\it gas}\cite{Ethereum2022:GAS}.
Gas is also used as a metric to represent the size of a block by summing up the gas required
for executing all transactions within that block.
Users who submit a transaction specify the amount of ETH they will pay as a fee per gas
(typically measured in units of Gwei\footnote{
The smallest unit of Ether is wei, which is $10^{-18}$ ETH, and Gwei is $10^{-9}$ ETH.
}).
This is referred to as the gas price.
The amount obtained by multiplying the gas price by the actual gas consumed becomes the gas fee.

After the London update in 2021, the gas fee was divided into {\it base fee} and {\it priority fee}.
The price of the base fee per gas is determined by the protocol based on the congestion of the
preceding blocks.
The maximum size of a block is 30 million gas, and validators creating a block aim to include
transactions up to half of this maximum size, which is 15 million gas.
If the actual size of the created block is larger than this benchmark, the base fee for the
subsequent block can be increased by up to 12.5\%.
Conversely, if the block size is smaller than the benchmark, the base fee for the following block
will be decreased.
On the other hand, the price of the priority fee is set at the discretion of the transaction submitter,
based on how much they want their transaction to be prioritized.

The majority of the gas fee, which is the base fee, is burned and removed from ETH circulation.
The priority fee becomes the income of the validator who proposed that block (before The Merge,
it became the income of the miner who produced the block).

This design is believed to be based on the following reasons\cite{Buterin2019:EIP1559}.

\begin{itemize}
\item The base fee can be automatically determined based on the protocol, making fee
prediction possible.
\item By burning the base fee and not passing it to validators, the following can be achieved:
\begin{itemize}
\item Enforce that fees for transaction execution can only be paid in ETH, establishing the economic value of ETH.
\item Suppress strategic actions\footnote{
For instance, unjustly raising the fees by giving a portion of the fees back to users as a kickback
or by injecting dummy transactions oneself.
} by validators concerning gas fees.
\item Counteract inflation\footnote{
Not limited to Ethereum, there is a naive understanding accepted in the so-called crypto community
regarding the relationship between the scarcity of currency and inflation.
}.
\end{itemize}
\end{itemize}

\paragraph{Rewards and Penalties}
By participating in the aforementioned tasks as a validator, rewards can be obtained,
and there are penalties for neglecting the work\cite{Ethereum2022:Rewards}\cite{Ethereum2022:ValidatorFAQ}.
Both are applied to the deposit on a per-epoch basis.

The reward each validator receives is proportional to their respective effective balance and
inversely proportional to the square root of the total deposit amount by all validators
(details will be explained later in section~\ref{sec-eth-pos-model}).
The effective balance is an integer value rounded in ETH units, with a maximum value of 32 ETH.

Rewards are paid in compensation for the following tasks:
\begin{itemize}
\item Timely attesting to the source checkpoint.
\item Timely attesting to the target checkpoint.
\item Timely attesting to the latest block (participating in the beacon committee).
\item Participating in the sync committee.
\item Proposing a block.
\end{itemize}
The reward decreases as the attestation is observed to be made at a later timing on a per-slot basis.
Furthermore, if sufficient attestations are not obtained, leading to a failure in justifying
or finalizing checkpoints, not only are penalties applied to validators on the minority side,
but rewards for attestations are also not paid to those on the majority side (to prevent acts
of sabotaging other validators through censorship or Denial-of-Service (DoS) attacks).

The block proposer receives additional rewards proportional to the number of valid attestations
included in the block.
Furthermore, they can also earn rewards by incorporating evidence of malicious behavior by
other validators into the proposed block.

On the other hand, penalties are imposed in the following cases, and are confiscated from
the participant's deposit (the basic amount of the penalty is equal to the reward that would
have been received had they participated correctly):
\begin{itemize}
\item When not responding after being selected for the sync committee.
\item When not attesting to a set of checkpoints.
\end{itemize}

When validators corresponding to more than $\frac{2}{3}$ of the total deposit are operating
normally, the penalty imposed on the minority side is not significant.
If the majority is less than $\frac{2}{3}$ (in cases where finalization does not occur),
the penalty becomes larger.

Furthermore, if the following actions are observed, the validator will be expelled ({\it slashed}).
Specifically, $\frac{1}{32}$ of the deposit amount is confiscated, and after a grace period
of 36 days, the participant automatically loses their qualification as a validator.

\begin{itemize}
\item Proposing multiple different blocks in the same slot.
\item Attempting to modify history.
\item Double voting.
\end{itemize}
When there are multiple validators subject to slashing, the amount confiscated from each
validator's deposit increases (as a countermeasure against collusion).

In summary, by participating in Ethereum as a validator, one can earn rewards in ETH at
an annual rate of $2\% \sim 20\%$\cite{Ethereum2022:ValidatorFAQ}.
Typically, it is less than $5\%$.

\paragraph{Exit}
A validator can be removed from the list in the following ways:

\begin{itemize}
\item When penalties cause the deposit to fall below the threshold (16 ETH) or
when the validator loses their qualifications due to slashing.
\item By declaring their intention to exit through a signature-required exit message.
\end{itemize}

The number of validators that can exit within a certain period is limited
(the number of validators that can join is also limited).
Once a validator exits, they cannot return with the same address.

\section{Concerns about the Sustainability of the Ethereum PoS}
\subsection{Before The Merge}
Past Ethereum was based on PoW, and fundamentally, it is believed that there was an equilibrium
relationship between the market value of ETH and the imposed electricity cost, as pointed out
by the authors in \cite{Iwamura2019:BitcoinMonetaryPolicy}\cite{Saito2019:Stable}.

In addition, since Ethereum serves as a platform for executing smart contracts, there exists
a mismatch in participation incentives as pointed out in \cite{Shudo2018:Portability}.
Specifically, for miners involved in maintaining the blockchain, the acquisition of ETH and
its price appreciation serve as incentives.
On the other hand, for application developers and users, the benefits derived from executing
smart contracts are the primary participation incentives.
As long as application developers and users have a desire to deploy and execute smart contracts,
it can be reasoned that they would support buying ETH to pay for gas fees.
However, an increase in gas prices or the price of ETH can negatively impact user
participation incentives.

Due to this mismatch in incentives, there was a risk that the Ethereum blockchain could halt
for reasons unrelated to applications, in the event of a sharp drop in the price of ETH\footnote{
In contrast, for the Bitcoin blockchain, the applications deployed on its foundation are
essentially only the transfers of its native token, BTC.
As initially thought, if the BTC transfer function is mainly used as a means of payment
and the transaction fee is at a level that cannot be overlooked by users, a mismatch in
participation incentives between miners and users could arise, similar to the aforementioned
Ethereum PoW.
However, in reality, since most transactions are made by users who view BTC as an
``asset (investment target)'' and utilize its transfer function expecting its price appreciation,
it can be said that the participation incentives of application users and foundation maintenance
participants (miners) currently align on the Bitcoin blockchain.
}.

However, in the case of PoW, the price of ETH was anchored, so to speak, to electricity costs,
tying it to the real economy.
Through the purchase and sale of hardware and the turning on and off of power, rapid entries
and exits of miners that are linked to the price of ETH are believed to have been suppressed.
Although in Ethereum's PoW, the use of GPUs was mainstream.
Given that the hardware is general-purpose, and with the increasing momentum for machine
learning applications today, there's high demand, making sales easier.
(This might be one reason why former miners could see The Merge, which could seem to
encourage their exit, as a realistic choice.)

\subsection{After The Merge}\label{sec-after-merge}
Under PoS introduced by The Merge, the primary cost borne by Ethereum validators is
no longer the electricity cost, given the significant reduction in computational resources
required for blockchain maintenance such as block creation.
Instead, it is thought to be the opportunity cost of ETH demanded as a deposit requirement
for validator qualification.
Specific opportunity costs include the gain or loss from selling ETH at any desired timing,
or the opportunity to earn ETH and other tokens at a higher return rate by depositing or
lending ETH to other blockchain platforms or DeFi.
However, as long as an increase in the price of ETH is expected, the cost of missed selling
opportunities is minimal, and the value of ETH earned as a reward for participation as
a validator is also expected to rise.
Therefore, the expected profit and loss for validators, that is, their participation incentive,
will depend on the price movement of ETH itself.

While the participation incentive for validators being tied to the price trend of the native
token is unchanged from PoW, a significant difference in PoS is the reduced value of what
can serve as an anchor in token price formation (linking the token price to a value in the real
economy, equivalent to the electricity cost in PoW).
On the other hand, for application developers and users, the profit from executing smart contracts
serves as a participation incentive, while the gas fees required for the execution of smart
contracts act as a disincentive.
This remains true regardless of whether it is PoW or PoS.

In other words, under Ethereum PoS, while an anchor linking to the real economy has been lost
within the components of validators' participation incentives, the {\it profit from executing smart contracts},
which is a component of the participation incentives for application developers and users, is
thought to serve as an anchor tying the ETH price to the real economy through its equilibrium
with gas fees in the long term.

Consequently, when considering the equilibrium price of ETH in Ethereum PoS, it is necessary
to think about the participation incentives of smart contract developers and users in Ethereum,
that is, the conditions under which they enter and exit Ethereum.

If Ethereum holds a monopolistic position as a platform for executing smart contracts,
there is a possibility that developers and users might be forced to give up using the platform
due to surging gas fees, and indeed, there have been several instances of such occurrences after
the year 2021 when prices often soared.
However, Ethereum is no longer the only platform for executing smart contracts.
Users can now compare the level of gas fees, convenience, and stability of various platforms and
choose which blockchain to use.
This trend has already begun, as observed in the NFT\cite{Entriken2018:NFT} market\cite{OpenSea:Web}
and other areas.
The platform for executing smart contracts is exposed to competition, and there is no guarantee
that Ethereum will not end up being a loser in this competition.

\subsection{Ethereum PoS Modeling}\label{sec-eth-pos-model}
Here, we model {\it utility and cost to users}, {\it gas market},
{\it validators' income and expenses}, and {\it changes in the total amount of ETH}
using simple mathematical representations and discuss the effects of competition
among multiple execution platforms.

\paragraph{Utility and Cost to Users}
Let the utility for user $x.u$ resulting from the execution of transaction $x$ be denoted as
$\mathcal{U}^{x.u}_x$ and be measured in fiat currency.
Also, the utility for $x.u$ from executing $x$ specifically on Ethereum can be represented
as $\mathcal{U}^{x.u}_{x_E}$ and is similarly measured in fiat currency.

Let the gas for transaction $x$ be denoted as $x.g$.
The base fee per gas $f_h$ in the block with block number (or {\it height}) $h$,
that is, the set of transactions $X_h$, is based on \cite{Buterin2019:EIP1559} and is
derived from the block's maximum size $G$.
The fee increase (or decrease) ratio is determined by the ratio of the actual block's gas to
the reference $\frac{1}{2}G$, using the function $F_\Delta$.
This relationship can be represented as:
\[
f_h = 
\begin{cases}
1\:\text{Gwei} & h = h_0\\
F_\Delta\left(\sum_{x \in X_{h-1}}x.g, G\right) f_{h-1} & h > h_0
\end{cases}
\]
where $h_0$ is the block height at the time of the London Hard Fork.

Let the priority fee per gas that the user $x.u$, the initiator of $x$, intends to pay
at block number $h$ be denoted as $p^{x.u}_h$.
If the exchange rate between ETH and legal currency (here, let us assume USD) is $\theta$,
then the gas fee for $x$ can be expressed as\footnote{
Incidentally, the ratio of $\sum_{x \in X_{h-1}}x.g$ to $\frac{1}{2}G$, which is a determinant
for $f_h$, indicates the situation where the demand and supply for transaction processing are
relaxed or constricted.
This can influence the priority fee per gas $p^{x.u}_h$.
It can be understood that when fluctuations in the base fee per gas, which are predetermined
(and automatically decided) in the protocol, are not sufficient for demand-supply adjustments,
the priority fee per gas set based on user discretion is utilized as a supplementary
demand-supply adjustment function.
}:
\[
(f_h + p^{x.u}_h)\:x.g\:\theta
\]

From here, the rational conditions for user $x.u$ to submit $x$ targeting the slot
corresponding to Ethereum's block height $h$ can be represented by the following equation:
\begin{equation}\label{equ-tx}
\mathcal{U}^{x.u}_{x_E} - (f_h + p^{x.u}_h)\:x.g\:\theta \geq 0
\end{equation}

\paragraph{}
That is, for user $x.u$, the greater the utility $\mathcal{U}^{x.u}_{x_E}$, and the
smaller the gas fee $(f_h + p^{x.u}_h)\:x.g\:\theta$, the stronger the incentive to
participate in Ethereum and execute $x$.

To increase the utility $\mathcal{U}^{x.u}_{x_E}$, one might consider offering features
on Ethereum that allow for the execution of more advanced smart contracts.
Indirect effects or those with external economic effects include the deployment of a
large and diverse range of applications.
This allows users to selectively use them, consistently enjoying high-quality services
at a low cost.
Furthermore, combining various applications could provide even more advanced services.
Additionally, the track record of chaining blocks that are costly to create, enhancing
their tamper resistance, is believed to foster a sense of trust in the stable operation
of the platform as an execution base.

On the other hand, to reduce the gas fee $(f_h + p^{x.u}_h)\:x.g\:\theta$, given that
$x.g$ is fixed for $x$, it would be desirable to reduce:
1) the gas price $(f_h + p^{x.u}_h)$, and/or
2) the exchange rate between ETH and the legal currency (USD), denoted by $\theta$.

\addtocounter{footnote}{-1}
Regarding 1), increasing the maximum block size $G$ can reduce the gas price (see the
previous footnote\footnotemark{}).
The {\it sharding} (horizontal distribution) that Ethereum plans to introduce in the future\footnote{
Instead of all validators verifying all transactions in a block, each validator equally handles
a partitioned management section ({\it shard}) within the block.
However, the load distribution of validator processing only increases $G$ by $m$-fold,
and in at least past designs, $m$ is set to 64.
} is likely aiming for such an effect.
Nonetheless, it is essential to note that if the size is not sufficiently large compared to
the demand\footnote{While not uncommon in economic events, there may also be a need to cope with
temporary and drastic fluctuations in demand.
}, one cannot adequately prevent the situation where exceeding the processing capacity
threshold causes a surge in gas prices.

\paragraph{Gas Market}
The discussions thus far have been based on the assumption that Ethereum is the only platform
for executing smart contracts.
However, as mentioned earlier, Ethereum is no longer the sole platform for smart contracts.
Therefore, we need to consider scenarios where application developers and users can choose
execution platforms other than Ethereum.

Let's consider a set of competing platforms for executing smart contracts, denoted as $L$
(for the sake of discussion, assume that all virtual machines across these platforms can
execute operations equivalent to $x$, and the cost for the required gas, $x.g$, can be
measured in their respective native tokens as well as in fiat currency).
For any platform $l \in L$, let its gas price be denoted as $f^l$ (we do not distinguish
between basic/priority fees here for simplicity, given that some platforms might not have
this differentiation), and the exchange rate between its native token and fiat currency
be $\theta^l$.
If the utility of user $x.u$ executing $x$ on platform $l$ is represented by $\mathcal{U}^{x.u}_{x_l}$,
then the rational condition for $x.u$ to continue using Ethereum is represented by the
logical conjunction of Equation (\ref{equ-tx}) and the following Equation (\ref{equ-competition}).

\begin{equation}\label{equ-competition}
\forall l \in L{:} \quad \mathcal{U}^{x.u}_{x_E} - (f_h + p^{x.u}_h)\:x.g\:\theta \quad
\geq \quad \mathcal{U}^{x.u}_{x_l} - f^l x.g\:\theta^l - \mathcal{EC}^{x.u}_{x_{E/l}}
\end{equation}

\paragraph{}
Here, $\mathcal{EC}^{x.u}_{x_{E/l}}$ represents the (positive) external economic effects
associated with using Ethereum, as mentioned earlier.
When a user considers moving from Ethereum to another competing execution platform $l$,
this represents the cost (or lost utility) that arises separate from the gas fees required
for individual transaction processing (the lock-in effect to Ethereum).
If a competing platform has a larger positive external economic effect than Ethereum,
it is conceivable that $\mathcal{EC}^{x.u}_{x_{E/l}}$ can take a negative value.

As mentioned earlier, for the sake of simplifying the discussion, we assume here that
transactions $x$ can be processed equally on Ethereum and other competing platforms
(providing the exact same service), so $\mathcal{U}^{x.u}_{x_E}$ and $\mathcal{U}^{x.u}_{x_l}$
are equal.
Furthermore, for simplification, if we assume the external economic effect of Ethereum,
$\mathcal{EC}^{x.u}_{x_{E/l}}$, to be 0 (the validity of which is discussed in
section~\ref{sec-discuss-churn}), equation (\ref{equ-competition}) will become a condition
based on the relationship between the gas fees (converted to fiat currency) in each
platform, i.e., the gas fee on Ethereum, $(f_h + p^{x.u}_h)\:x.g\:\theta$, and the gas fee on
the competing platform, $f^l x.g\:\theta^l$.
Given that $x$ represents the same computation, $x.g$ is the same regardless of the platform.
Thus, equation (\ref{equ-competition}) boils down to the relationship between
$(f_h + p^{x.u}_h)\:\theta$ and $f^l\:\theta^l$, i.e., the relationship between gas prices
on each platform, $(f_h + p^{x.u}_h)$ and $f^l$, and the ratio of the native token to fiat
currency, $\theta$ and $\theta^l$.

This competition fundamentally forms a market for gas.
Specifically, gas, which represents the {\it amount of code execution $+$ data storage} that
can be verified for its authenticity, is treated as a commodity.
Ethereum (layers 1 and 2)\footnote{
The layer 1 refers to the foundation of Ethereum itself as described in this paper, while the
layer 2 refers to the general term for technologies that execute transactions off-chain from
the blockchain's perspective and make their proofs verifiable on the layer 1.
} competes with several other smart contract execution platforms based on its supply capacity
and price in this market.

In the current market, where Ethereum users overwhelmingly dominate among smart contract
execution platforms, it's often the case that $\theta^l < \theta$.
Therefore, even if the gas price of Ethereum $(f_h + p^{x.u}_h)$ and that of competing platforms
$f^l$ are at the same level, the gas fee in terms of fiat currency would be lower on the
competing platforms.
As a result, as we already see instances of platforms like Polygon\cite{Polygon:web} being used
in markets such as NFTs, non-Ethereum platforms may become more preferred.
If the ETH held by $x.u$ is sold, then $\theta$ would decrease, moving towards equilibrium
($\theta^l = \theta$, and consequently, $(f_h + p^{x.u}_h)\:x.g\:\theta = f^l x.g\:\theta^l$).

It is worth noting though that Ethereum is not unresponsive to issues such as the surge in
gas prices and the delay in transaction processing.
They are attempting to increase processing capabilities with effects similar to expanding
the maximum block size through the previously mentioned sharding,
and also aiming to incorporate layer 2 technologies such as {\it ZK-rollup}\cite{Ethereum2023:zkRollup}.

Rollup is a mechanism for executing transactions offline, while ZK stands for {\it Zero Knowledge}.
ZK-rollup involves executing transactions on the layer 2 (seen as {\it offline} from the layer 1) and
verifying their correctness on the layer 1 without revealing the contents of the transactions.
Given its capability to execute and verify batches of transactions, it is anticipated to
address scalability challenges.
Compared to the conventional method of processing all transactions on the layer 1, under the
ZK-rollup mechanism, the processing volume on the layer 1 can be reduced, for instance, to
about 0.3\%\cite{Loopring:web}.
Even if the gas price or the fiat currency price of ETH that layer 2 services should pay to
the layer 1 remains the same, the fees users pay in, say, USD on layer 2 services could decrease
to about $\frac{1}{300}$ of the traditional costs.
As a result, the amount of ETH users need to hold for executing smart contracts will decrease,
potentially exerting downward pressure on the fiat currency price of ETH.
However, if the reduction in per-transaction fees attracts more demand and the Ethereum
ecosystem, including the layer 2, possesses the capacity to handle this increased demand, it might
not only retain its advantage over other competing platforms but also potentially maintain
the fiat currency price of ETH at an appropriate level.

From the above, it can be said that Ethereum's aim for scaling through sharding and the layer 2
after The Merge holds significance as a strategy to maintain its raison d'être.

\paragraph{Validators’ Income and Expenses}
In a given epoch $e$, suppose there are $n$ validators in operation.
Let the effective balance of the $i$-th validator be denoted as $b^i_e$, and based on
its behavior, two types of coefficients are prepared according to the rewards and
penalties system\cite{Ethereum2022:Rewards}.
\begin{itemize}
\item A coefficient $r^i_e$ determined solely by the actions of the validator $i$ and
independent of $n$.
\item A coefficient $w^i_e$ determined by the block proposals and participation
in sync committees, which has probabilistic occurrences based on $n$ for each slot.
\end{itemize}
For the $s$-th slot in $e$, namely at height $h = F_B(e, s)$ (where $F_B$ is a function
to derive the block number from the epoch and slot, and can return $\bot$ to represent
absence) and the block $X_h$ (a set of transactions), the total priority fee can be
expressed as
\[
P_{e,s} = 
\begin{cases}
0\:\text{wei}& h = \bot\\
\sum_{x \in X_h}p^{x.u}_h\:x.g& h \neq \bot
\end{cases}
\]
Then, the expected income of the validator $i$ in $e$ can be approximated by
\[
\left(\frac{r^i_e\:b^i_e}{\sqrt{\sum_{j=1}^n{b^j_e}}} + \frac{1}{n}\sum_{s=1}^{32}\left(P_{e,s} + \frac{n w^i_e\:b^i_e}{\sqrt{\sum_{j=1}^n{b^j_e}}}\right)\right)\theta
\]

For simplification, assume that the effective balance of all validators is always
32 ETH, implying that every validator always performs their duties flawlessly.
In this scenario, let the reward coefficient determined solely by the validator's
behavior be $R$, the reward coefficient for probabilistic opportunities occurring in each
slot based on $n$ be $W$, and the average priority fee per slot be $P$.
Then the expected income per validator in a given epoch can be expressed as:
\[\frac{4R\sqrt{2n} + 32(P + 4W\sqrt{2n})}{n}\theta\]

Let $\alpha^i$ represent the cost for validator $i$ to stay for a new epoch, considering
the electricity costs required for computer verification work and the opportunity costs
associated with the deposit.
Assuming this is measured in fiat currency, the rational condition for $i$ to remain as
a validator (or to join anew) for that epoch can be represented by the following equation:
\begin{equation}
\frac{4R\sqrt{2n} + 32(P + \revDeleted{\frac{1}{n}}4W\sqrt{2n})}{n}\theta\ \geq \alpha^i
\end{equation}

\paragraph{}
It is believed that the dominant factor in $\alpha^i$ is the opportunity cost associated
with the 32 ETH deposit.
This could potentially serve as a motivation to exit, especially when $\theta$ is decreasing,
with validators wanting to minimize losses.
The left-hand side is inversely proportional to $n$, so the exit of other validators is
welcomed (cf. {\it discouragement attacks}\cite{Buterin2018:Discouragement}), and the incentive
to join decreases as $n$ grows larger.
Even if exits of other validators are welcomed, it is not guaranteed that the remaining
validators will have a greater incentive to stay when $\theta$ is decreasing.
Based on this, there is room for external discouragement attacks aimed at halting the
blockchain in practice, specifically by harming validator profits to encourage their exit.

\paragraph{Changes in the Total Amount of ETH}
Based on the notation above, during epoch $e$, the total ETH amount corresponding to the
sum of the base fees given by
\[
\sum_{s=1}^{32} f_h\sum_{x \in X_h} x.g
\]
is burnt
(for simplicity, we assume for all $h = F_B(e, s)$, $h \neq \bot$).

The total income that all validators can obtain in a given epoch is the expected
income per validator multiplied by $n$.
Of this, the amount related to $P$ represents a transfer of existing ETH from the
users submitting transactions to the validators proposing blocks.
If we exclude this amount, the increase (or decrease) in the total amount of ETH,
denoted $\Delta_e$, in epoch $e$ can be expressed by the following equation.
\begin{equation}\label{equ-delta}
\Delta_e = 4\sqrt{2n}(R + \revDeleted{\frac{1}{n}}32W) - \sum_{s=1}^{32} f_h\sum_{x \in X_h} x.g
\end{equation}

\paragraph{}
Within the Ethereum community (or the broader so-called crypto space), a naive
understanding of the relationship between currency scarcity and inflation is widely
accepted.
Therefore, if $\Delta_e$ is positive, ETH holders might perceive that the value of
their holdings has been diluted.

The first term on the right-hand side is in the form $\sqrt{n}$, which, as $n$ increases,
incrementally boosts the total amount of ETH, albeit at a decreasing rate.
Thus, the community would not welcome an excessive increase in $n$.
Additionally, an excessive decrease in $n$ is associated with a decrease in $\theta$, and
therefore would not be welcomed either.

The increase in the base fee of gas (the second term on the right-hand side) is welcomed
by the community because it can lead to an increase in the burning of ETH and consequently,
a decrease in the total amount of ETH.
However, if the increase in the base fee is due to a rise in its unit price\footnote{
Since $f_h$ is subject to positive feedback from $\sum_{x \in X_{h-1}} x.g$, it can be said
that this is the case unless $G$ becomes larger.
}, it implies that blocks are congested, and the profits of $x.u$ are compromised
(a mismatch in incentives).

Using the layer 2 increases the profit of $x.u$.
However, if there is not a sufficient increase in usage to match the decrease in gas prices,
there is a possibility that $\theta$ will decrease.
Furthermore, by alleviating block congestion, the unit price of the base fee might decrease.
This could lead to a reduction in the burning of ETH funded by the base fee and,
consequently, an increase in the total amount of ETH.
In such a scenario, the community might not welcome it, thinking that the value of ETH has
been diluted (a mismatch in incentives).

\subsection{Potential Ethereum PoS Outage}\label{sec-eth-pos-stop}
As discussed in section~\ref{sec-after-merge} and confirmed in the modeling in
section~\ref{sec-eth-pos-model}, the structure of incentive mismatch between validators and
users persists in Ethereum PoS.
In Ethereum PoS, for validators involved in the maintenance of the blockchain, both the
acquisition of ETH and its price appreciation are incentives.
On the other hand, for users, as long as they have to pay the gas fees required for executing
transactions in ETH, an increase in the price of ETH diminishes the incentive to participate
in Ethereum.
Given the current scenario where other execution platforms are emerging alongside Ethereum,
users have the real option to switch to competing platforms.
That is to say, considering the continued presence of incentive mismatch between validators
and users in Ethereum PoS, and the shift from an environment where Ethereum was almost the
sole execution platform to a more competitive situation, it is not necessarily true that the
transition to PoS has made Ethereum more stable or secure.

In Ethereum PoW, there was a risk that a significant drop in ETH's price could erode trust
or even halt the blockchain.
It remains to be seen if this risk has been mitigated with the transition to PoS.

As previously mentioned, in Ethereum PoS, for validators involved in the maintenance of the
blockchain, both the acquisition of ETH and its price appreciation are incentives.
Therefore, when the price rises, new entrants come in, and when it falls, they exit.
However, in Ethereum PoS, it is believed that sudden entries and exits are curbed by the fact
that the expected reward is inversely proportional to the square root of the total deposit
amount, and by limiting the number of validators that can enter or exit within a certain period.
Moreover, in PoW, when the price of ETH declined, miners could choose to turn off their hardware
to save on costs.
However, in PoS, there is a penalty for going offline, so turning off the system and exiting
could actually incur costs.
Given this, the Ethereum community, including its designers, likely views Ethereum PoS as having
a superior mechanism to deter exits compared to PoW.

Despite the implementation of such measures to deter exits, there could be potential {\it loopholes},
such as selling the rights of a PoS validator to someone else to effectively exit the system.
If a buyer is found, one could simply sell their private key to transfer their validator rights,
effectively allowing them to withdraw.
Such a trade merely changes the entity acting as a validator, so when viewed solely within the
confines of Ethereum, it seems to have no impact on the price or trustworthiness of ETH.
However, the scenario changes if there are entities wishing to devalue Ethereum.
By purchasing the rights of a validator and then deliberately failing to fulfill the
responsibilities of a validator (e.g., taking the validator's computer terminal offline), they
could potentially bring Ethereum to a halt in the worst-case scenario.

If the price of ETH were to plummet, the cost of penalties for validators going offline
(failing to fulfill their responsibilities) would decrease, making the simple option of
turning off their power and exiting more feasible.
Unlike exits due to slashing, validators going offline due to a drop in ETH's price could
affect both majority and minority validators equally.
This increases the likelihood that more than $\frac{1}{3}$ of the total validators' deposits
could go offline, making finality harder to achieve.
This could damage Ethereum's credibility as a platform, which in turn could lead to a further
decline in the price of ETH (potentially triggering a negative spiral).

From these observations, it can be inferred that in Ethereum PoS, a decline or crash in the
price of ETH could still lead to a loss of trust in Ethereum or a halt in the blockchain.
It is not definitive that the mechanisms in place to prevent such events are stronger compared
to PoW.

\section{Discussion}

\subsection{Stability of ETH Total Volume and Monetary Value}
A concern regarding Ethereum PoS, briefly mentioned in the discussion related to
equation (\ref{equ-delta}), is the seeming misconception embedded in the rules that
the quantitative stability of ETH supply, such as the burning of base fees for transaction
execution, leads to the stability of its currency value.
However, we had already pointed out that a mere fixed supply in crypto-assets should bring
about not stability, but instability in its market price\cite{Iwamura2014:BitcoinMonetaryPolicy}.
This concern has been manifested in the extremely volatile price movements of many crypto-assets,
including bitcoin.

Indeed, ETH is different from bitcoin.
While bitcoin mainly serves as a speculative asset, beyond its use as a means of payment,
if one can grasp its market price moment by moment, it can still function as a
settlement currency even without value stability.
On the other hand, Ethereum is designed to act as an execution platform for contracts
under the guise of so-called smart contracts.
Therefore, for ETH, which serves as the unit for contract representation and performance,
i.e., a measure of value, the importance of currency value stability (predictability) should
be far greater than that for bitcoin.

In the case of bitcoin, one could argue that the very unpredictability of its monetary value
forms part of its allure as a speculative asset.
However, for ETH, the instability in its formation of monetary value, when compared to fiat
currencies like the dollar, euro, or yen, which currently shoulder the display and execution
of many contracts worldwide, we believe poses a significant drawback as a measure of contract value.
The demand for ETH thus far seems not only driven by the need for a means of payment and measure
of value for smart contracts on Ethereum but also by its speculative price movements.
If Ethereum, transitioning to PoS, aims to shift its primary purpose from the speculation
generated by PoW to a foundation for smart contracts, then the misconception that a quantitative
stability in ETH supply leads to stability in currency value might hinder Ethereum's future
development.

For ETH, it might be more straightforward and beneficial to seek its value foundation in the
utility of services as a platform for executing smart contracts.
This approach could not only lead to stability in its value as a currency but also contribute
to its efficiency as a foundation for executing smart contracts.

\subsection{Monopoly and Competition}\label{sec-mono-comp}
Our concern is whether Ethereum, in planning and executing its transition to PoS, truly
recognized the depth of its current situation.
Up to this point, Ethereum has achieved a position that could be described as almost
monopolistic as the standard platform for executing so-called smart contracts.
However, with the emergence of competing platforms, the possibility of a shift to a
competitive environment that could threaten this position cannot be denied.

In the current PoS, the condition for participating as a validator is to deposit ETH.
This is likely based on the idea that holding ETH as a cryptographic asset should
naturally serve as an incentive to maintain its value.
We do not intend to dispute this understanding.
However, what needs to be noted is that even entities that currently hold ETH as a
cryptographic asset may, depending on their situation as ETH-based contract parties,
have an incentive not just to maintain but to collapse its value.

When might such a situation occur? The answer is simple.
If the validators are debtors, not creditors, in ETH-based contracts, a decrease in
the real value of ETH would benefit them.
Therefore, if the majority of validators:
1) have balance sheets that possess both ETH-based debts and real assets such as other
currency-denominated receivables, stocks, or real estate, or
2) are entities deeply committed to competing execution platforms,
then such a group of validators might have a greater incentive to depreciate, rather
than maintain, the value of ETH as a whole.
(Even in the former case, if there are competing execution platforms and switching to them
is easy, they might not face much criticism from users even if they were to force Ethereum
to halt.)
By borrowing 32 ETH from others and becoming a validator, one can easily establish the
position mentioned above.

Nevertheless, the incentive to depreciate the value of ETH might not blatantly manifest in
the actions of the validator group.
Actions by validators to impair the value of ETH could be profitable in the short term for
those engaging in them, but might not benefit them in the longer term.
However, our concern is that the current PoS rules do not exclude the unfortunate or
malicious possibility that a majority of validators could be overwhelmed by excessive debt
denominated in ETH.
This seems fundamentally contradictory to Ethereum's current direction, which aims to support
the development of ETH-denominated credit and debt transactions as a so-called smart contract
execution platform.

Of course, the deliberate value impairment issue among validators could also occur in PoW.
However, we should not overlook the fact that many cryptocurrencies, including bitcoin,
have experienced such volatile value fluctuations that, while they have become subjects of
speculation, they have not been able to serve the role of supporting contractual relationships
as a measure of value as we call it in monetary theory.
One of the reasons the so-called 51\% attack, continuously discussed in BTC, has not actually
occurred up to now is nothing more than a kind of fortune created by the {\it flaw} of price
instability brought about by the inadequacy of the target adjustment algorithm in BTC,
a cryptocurrency based on PoW\cite{Iwamura2019:BitcoinMonetaryPolicy}.

Is there a way to eliminate the incentive for currency value managers to impair the value
of their own currency?

The first method is to strictly set eligibility criteria for those who act as currency value
managers and place them under some form of ``public'' oversight.
The demand for {\it central bank independence} in the world of fiat currency arises from the
recognition that the government, which establishes the central bank, is often a major holder
of debt denominated in the fiat currency.
As a result, it tends to have an incentive, at least in the short term, to impair the value
of its national currency, leading to an induced inflation.

The second method is {\it competition} among currency value managers.
The theory of competitive currency issuance developed by Friedrich A. Hayek suggested that
if the choice of which currency to use is left to the individual's freedom rather than
enforced by law, at least in the long run, the incentive to maintain currency value would
likely prevail over the incentive to impair it.
The authors believe that the validity of his argument was at least indirectly tested in the
form of the global end of inflation brought about by the competition to maintain currency
value among central banks after the collapse of the fixed exchange rate system.
Drawing from that experience, perhaps competition among various smart contract execution
platforms is the desirable future.

\subsection{Is It Realistic to Switch Contract Execution Platform?}\label{sec-discuss-churn}
The network externality inherent in computer software poses a barrier to free competition.
In section~\ref{sec-eth-pos-model}, was it reasonable to proceed with the discussion assuming
the external economic effect of Ethereum, denoted as $\mathcal{EC}^{x.u}_{x_{E/l}}$, to be $0$?
Especially for smart contracts, such as DEX (Decentralized Exchange), which exchange fungible
tokens with alternatives based on certain rates, it is believed that network externality has
a significant impact.
However, as pointed out multiple times in this paper, this is not the case for non-fungible tokens
(NFTs) and the like, which lack substitutability.

When deploying a new smart contract independent of past ones, switching is not an issue.
Even when upgrading a smart contract on Ethereum, once deployed, it cannot be modified
(though it can be deactivated), so there is a practice of re-deploying it from scratch.
In this case too, switching rarely becomes a problem.

Transferring data from an old contract to a new one can be more challenging, but methods
have been proposed, for instance, in \cite{Shudo2018:Portability}.
Furthermore, in the case of Ethereum layer 1, all transactions are recorded, so it is not
difficult to reproduce the latest state offline (and third parties can verify its accuracy).
This has to be ported to the destination platform, but if there are platforms that see
migration from Ethereum as a business opportunity, they will likely provide mechanisms to do
this at a low cost.
Service providers specializing in facilitating these switches might also emerge.

\subsection{Factors that Cause Platform Switching}
The level of gas fees is not the only factor for switching platforms.
If Ethereum is perceived to lack stability, a switch might occur.
For instance, as illustrated in section~\ref{sec-eth-pos-stop}, stability could be
compromised due to attacks from competitors.

Furthermore, compared to the cost of maintaining control over more than
$\frac{1}{3}$ of the total deposit amount in ETH in the long term
(since they would be forced to exit if they either testify to prevent finality
or neglect to testify, necessitating a continuous deposit), if a greater value
resides on Ethereum's smart contracts, as already discussed in section~\ref{sec-mono-comp},
entities that could profit by invalidating them or resolve liabilities might have a
rational incentive to halt Ethereum by launching attacks that continuously prevent
checkpoint finality.

\section{Conclusions}
In this paper, we introduced the perspectives of ETH price levels and competition among
smart contract execution platforms into the discussion on Ethereum's sustainability.
We illustrated that there may be scenarios where Ethereum PoS cannot sustain when the
price of ETH falls.

In the study \cite{Cook2022:PosAttacks}, various potential attacks on Ethereum PoS are
discussed.
It argues that the security of PoS relies on the assumption that benevolent participants
hold a majority based on their deposit amounts.
If this condition is violated, it is technically impossible to fend off attacks, and
ultimately, the malevolent participants must be expelled by the power of the benevolent
community.
However, the study does not mention the price levels of ETH.
Can we expect the strength of the benevolent community if the price of ETH is declining?

A mismatch in incentives can lead to a situation where the underlying platform loses
trust, rendering applications inoperable.
There is a need for a new smart contract platform design that aligns the incentives
of users who develop and run applications with those who maintain the platform.

If the issues raised in this paper lead to a reconsideration of the widely adopted
Proof of Stake design or to the design of a new smart contract platform,
it would be fortunate.

\section*{Acknowledgment}
This work was supported by JSPS KAKENHI Grant Number JP21H04872.

\bibliographystyle{plain}
\bibliography{ethereum-pos-problem}

\end{document}